\documentclass[amsmath,amssymb,aps,prl,reprint,showpacs,twocolumn,floats,superscriptaddress]{revtex4-1}

\usepackage{graphicx}
\usepackage{amssymb}
\usepackage{amsmath}
\usepackage{hyperref}
\usepackage{footnote}
\usepackage{color}
\usepackage{multirow}
\usepackage{enumerate}
\usepackage{capt-of}
\usepackage{soul}
\usepackage[FIGTOPCAP]{subfigure}
\begin{document}
\title{High-fidelity dissipative engineering using parametric interactions}
\author{E. Doucet}
\email{emery\_doucet@student.uml.edu}
\affiliation{Department of Physics and Applied Physics, University of Massachusetts, Lowell, MA 01854, USA}
\author{F. Reiter}
\affiliation{Harvard University, Department of Physics, 17 Oxford St, Cambridge, MA 02138, USA}
\author{L. Ranzani}
\affiliation{Raytheon BBN Technologies, Cambridge, Massachusetts 02138, USA}
\author{A. Kamal}
\email{archana\_kamal@uml.edu}
\affiliation{Department of Physics and Applied Physics, University of Massachusetts, Lowell, MA 01854, USA}
\affiliation{Research Laboratory of Electronics, Massachusetts Institute of Technology, Cambridge, MA 02139, USA}

\begin{abstract}
Established methods for dissipative state preparation typically rely on resolving resonances, limiting the target state fidelity due to competition between the stabilization mechanism and uncontrolled dissipation. We propose a protocol devoid of such constraints, using parametric couplings to engineer dissipation for preparation of any maximally entangled two-qubit state. Our scheme allows high-fidelity entanglement generation with short convergence time, continuous control of the target state in the stabilized manifold, and is realizable with state-of-the-art superconducting qubit technology.
\end{abstract}
\maketitle
%
%
%
Quantum state preparation and preservation are cornerstones of any quantum information platform.
Standard methods of state preparation involve a set of unitary operations (or gates) on individual and multiple qubits to achieve a desired entangled state of the system. Such methods typically require multiple tightly synchronized pulse sequences and ancilla qubits, and complex algorithms to avoid unwanted interactions and contain the quantum information within the desired subspace. The prepared state also remains sensitive to environmental decoherence, which reduces the pure quantum state into a classical mixture and ultimately limits the power to harness quantum effects. 
%
%
Recent years have seen an emergence of an alternative approach embracing the environment instead of hiding it \cite{Kraus2008,Verstraete2009}. The basic idea relies on engineering suitable interactions with the environment that steer the reduced system towards a desired target state \cite{Plenio1999,Benatti2003}.  Besides engineering states inherently robust to dissipation, such dissipative state preparation precludes the need for any active control of the system and is relatively immune to initialization errors. Also, certain quantum operations become feasible only through dissipation; for instance, non-local interactions are essential for the stabilization of multi-partite entangled states such as the 3-qubit GHZ state \cite{TicozziViola2014,Reiter2016}. This has led to interest in dissipative preparation techniques being adopted in a gamut of quantum information platforms such as cavity QED \cite{Clark2003, Parkins2006}, trapped ions \cite{Lin2013,Schindler2013,Kienzler2015}, superconducting qubits \cite{Shankar2013,Schwartz2016}, Rydberg atoms \cite{Carr2013,Rao2013}, atomic ensembles \cite{Krauter2011}, and NV centers \cite{Li2012}.
\par
In this Letter, we present a novel paradigm for engineering dissipation using parametric driving. We demonstrate that this platform enables high-fidelity entanglement generation and control in a circuit-QED setup \cite{Blais2004}. In contrast to usual dissipation-engineering schemes, which rely on resonant driving, our scheme exhibits no tradeoff between fidelity and speed of stabilization protocol. It also leverages several technical and operational advantages offered by parametric driving, such as easy implementation relying on continuous-wave (CW) drives alone, strong coupling even in the non-resonant regimes unlike the strong-dispersive circuit-QED \cite{Schuster2007}, and in-situ tunability of parametrically-mediated interactions. The framework of parametrically-engineered dissipation presented here constitutes a significant addition to the parametric quantum toolbox, which is being wielded in burgeoning applications, such as entangling gates \cite{Allman2010,McKay2016,Reagor2018}, dynamical correction of qubit errors \cite{Lu2017}, nonreciprocal scattering \cite{Kamal2011} and synthetic magnetic fields \cite{Roushan2016}, holonomic gates for continuous-variable quantum information \cite{Mirrahimi2014} and even quantum annealing \cite{Puri2017}.
%
\par
\begin{figure}[t!]
\begin{minipage}[t]{\columnwidth}
\centering
\includegraphics[width=\columnwidth]{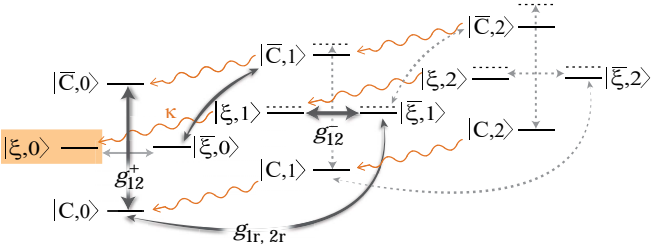}
\captionof{figure}{Coupling diagram for two-qubit entanglement stabilization. Thick black arrows denote resonant interactions, while dotted gray arrows denote off-resonant couplings. Wavy arrows depict photon decay from the resonator. The target state (highlighted with the colored box) and the complementary state (phase-flipped but same parity) are denoted with Greek letters $|\xi\rangle, |\bar{\xi}\rangle$ respectively. The states in the opposite parity manifold are denoted with $|C\rangle, |\bar{C}\rangle$ respectively.}
\label{Fig:Mechanism}
\end{minipage}%
\vspace{10pt}
\begin{minipage}[t]{\columnwidth}
\begin{tabular}{c|c|cccc|c}
$\sqrt{2}|\xi\rangle$ & $|C\rangle$ &  $\phi_{12}^{+}$ & $\phi_{12}^{-}$ & $\phi_{1r}$ & $\phi_{2r}$ & $c_{p}$\\ \hline\hline
$|gg\rangle - e^{i\psi}|ee\rangle$ & $|eg\rangle$ & $\pi/2 - \psi$ & 0 & 0 & $\psi$ & $\sigma_{1} + e^{i\psi}\sigma_{2}^{\dagger}$\\
$|ge\rangle - e^{i\psi}|eg\rangle$ & $|gg\rangle$ & 0 & $\pi/2 - \psi$ & 0 & $\psi$ & $\sigma_{1} + e^{i\psi}\sigma_{2}$\\
\end{tabular}
\end{minipage}
\captionof{table}{State vectors, pump phases, and respective parametrically-engineered collapse operators for arbitrary maximally-entangled states of even or odd parity.}
\label{Table:Bell}%
%
\end{figure}

\emph{Parametrically-engineered dissipation}:
%
Our proposed scheme for quantum state engineering consists of a system of two qubits and a single mode cavity (resonator) sharing a tunable coupler. The system is described by a generic Hamiltonian of the form
\begin{eqnarray*}
\mathcal{H} = \omega_{r} a^{\dagger} a + \sum_{j=1}^{2}\Big(\frac{\omega_{j}}{2}Z_{j}+ \tilde{g}_{jr}(t) X_j X_r \Big) + \tilde{g}_{12}(t) X_1 X_2,
\label{Eq:Hamil}
\end{eqnarray*}
where $X_{j} = \sigma_{j}+\sigma_{j}^{\dagger}; \; X_{r} = (a+ a^{\dagger})$. Each of the pairwise couplings, in addition to the usual static dispersive coupling, are parametrically modulated via a continuous wave pump as 
$\tilde{g}_{jk} (t) = \bar{g}_{jk} + g_{jk}^{+}\cos(\omega_{jk}^{+}t + \phi_{jk}^{+}) + g_{jk}^{-}\cos(\omega_{jk}^{-}t + \phi_{jk}^{-})$.
%
%
Different coupling terms can be activated in the circuit Hamiltonian by choosing the modulation frequencies as either the sum $\omega_{jk}^{+} = \omega_{j} + \omega_{k} $ or difference $\omega_{jk}^{-} = \omega_{j} - \omega_{k} $ of the resonance frequencies of the qubits and resonator. Figure \ref{Fig:Mechanism} depicts the generic coupling structure in the combined qubit-resonator manifold realized using the parametric pumps, truncated at the first three resonator levels. The qubit-qubit drives couple the states of same parity, while the qubit-resonator drives mix the two parity manifolds. The combined action of these drives is to recirculate population from any arbitrary two-qubit state with zero photons in the resonator,  into the state $|\xi,1\rangle$ i.e. the target qubit state with a single photon in the resonator --- this state then ultimately condenses into the target state $|\xi,0\rangle$ via the photon loss from the resonator. 
\par
Depending on the choice of drive frequencies, the scheme is capable of stabilizing any maximally-entangled two-qubit state with a given parity. For instance, for preparing even-parity states, the qubit-qubit interactions are modulated at frequencies $\omega_{12}^{+} =\omega_{1} + \omega_{2}  -2\chi_{1} -2\chi_{2}$, $\omega_{12}^{-} = \omega_{1} - \omega_{2}$, while qubit-resonator interactions are modulated at the four  frequencies $\omega_{1r}^{-} = \omega_{1} - \omega_{r} - \chi_{1} \pm \chi_{2}$ and $\omega_{2r}^{+} = \omega_{2} + \omega_{r} - \chi_{2} \pm \chi_{1}$. Here $\chi_{j} = \bar{g}_{jr}^{2}/(\omega_{r} -\omega_{j})$ represent the dispersive shifts induced on the qubit states due to respective static qubit-resonator coupling strength $\bar{g}_{jr}$. In a rotating frame defined with respect to the dispersive qubit-resonator Hamiltonian, this leads to an effective Hamiltonian of the form \cite{Supplement},
\begin{eqnarray}
\mathcal{H}_{\rm eff}^{\rm even} &=& g_{12}^{-} e^{i\phi_{12}^{-}}(1 - \langle a^{\dagger} a \rangle)\sigma_{1} \sigma_{2}^{\dagger} + g_{12}^+ e^{i\phi_{12}^{+}}\langle a^{\dagger} a \rangle \sigma_{1} \sigma_{2} \nonumber\\
& & + \; (g_{1r}^{-}e^{i\phi_{1r}^{-}}\sigma_{1} + g_{2r}^{+}e^{i\phi_{2r}^{+}}\sigma_{2}^{\dagger}) a^{\dagger} + h.c.
\label{Eq:EvenBell}
\end{eqnarray}
in the subspace truncated at lowest two levels of the resonator. Similarly, for preparing odd-parity states, choosing $\omega_{12}^{+} = \omega_{1} + \omega_{2}$, $\omega_{12}^{-} = \omega_{1} - \omega_{2} - 2\chi_{1} + 2\chi_{2}$,  $\omega_{1r}^{+} = \omega_{1} + \omega_{r} -\chi_{1} \pm \chi_{2}$, $\omega_{2r}^{+} = \omega_{2} + \omega_{r} -\chi_{2} \pm \chi_{1}$, leads to an interaction of the form
\begin{eqnarray}
\mathcal{H}_{\rm eff}^{\rm odd} &=& g_{12}^{+}e^{i\phi_{12}^{+}}(1 - \langle a^{\dagger} a \rangle)\sigma_{1} \sigma_{2} + g_{12}^{-}e^{i\phi_{12}^{-}}\langle a^{\dagger} a \rangle \sigma_{1} \sigma_{2}^{\dagger} \nonumber\\
& & + \; (g_{1r}^{+}e^{i\phi_{1r}^{+}}\sigma_{1}^{\dagger} + g_{2r}^{+}e^{i\phi_{2r}^{+}}\sigma_{2}^{\dagger}) a^{\dagger} + h.c.
\label{Eq:OddBell}
\end{eqnarray}
\par
It is worthwhile to highlight here the features of the Hamiltonians presented in Eqs. (\ref{Eq:EvenBell})-(\ref{Eq:OddBell}) that distinguish them from typical state preparation protocols. (i) As described above, the form of each of the coupling terms is uniquely set by the choice of frequency of the pump mediating the coupling. (ii) Additionally, the dispersive qubit-resonator couplings lead to a selective activation of the qubit-qubit couplings contingent on the photon occupation in the resonator. (iii) By virtue of the couplings being parametric, their strengths and phases are tunable through the amplitudes and phases of the pumps. This freedom of coupling parameters, combined with a decay on the resonator, provides a convenient method for realizing different parametrically-engineered collapse operators $c_{p}$. This allows the stabilization of distinct target states, i.e. $c_p|\xi\rangle=0$ \cite{Kraus2008}, through the choice of $\omega_{jk}$ and $\phi_{jk}$.  
\par
%
%
\par
\emph{Stabilization Mechanism}: 
%
We now describe how the above functionalities may be exploited to dissipatively stabilize any maximally-entangled two-qubit state. The generic coupling map in Fig. \ref{Fig:Mechanism} describes the stabilization mechanism for all four Bell states, with an appropriate relabeling of states and optimal choice of pump phases as stated in Table \ref{Table:Bell}. For the purposes of illustration, we describe the mechanism for the specific case of even-parity Bell state implemented using the effective interaction Hamiltonian of Eq.~(\ref{Eq:EvenBell}). Assume that the two qubits have been stabilized into $|\Phi_{-}\rangle = (|gg\rangle - |ee\rangle)/\sqrt{2}$. Under a bit flip, the parity of the state changes leading to a jump to the odd parity manifold spanned by $\{|ge\rangle, |eg\rangle\}\otimes|0\rangle$. Each of these odd-parity states is coherently coupled to the even-parity state $|\Phi_{+},1\rangle$, either directly using the qubit-resonator drives (for $|eg\rangle$) or through a combination of qubit-qubit and qubit-resonator drives (for $|ge\rangle$), while exciting the resonator. The remaining qubit-qubit drive $g_{12}^{-}$ pumps the population from $|\Phi_{+},1\rangle$ to $|\Phi_{-},1\rangle$, which then decays down to the target state $|\Phi_{-},0\rangle$ as the resonator loses the photons at rate $\kappa$. Any population that enters the state $|\Phi_{-},0\rangle$ remains unaffected by the qubit-qubit drive and the qubit-resonator drive, since $|\Phi_{-}\rangle$ is a dark state of both the qubit Hamiltonian and the engineered collapse operator $c_{p} = \sigma_{1} + \sigma_{2}^{\dagger}$. A phase flip on a qubit moves the population to $|\Phi_{+}\rangle$;  this state strongly couples to the odd-parity state $|ge, 1\rangle$ through the qubit-resonator interaction. The excitation of the resonator induces a dispersive shift of the qubit frequency making the $g_{12}^{+}$ drive off-resonant for this state; hence the dominant exit channel is the resonator decay which brings back the population to odd-parity manifold with no photons in the resonator. This is then repumped to $|\Phi_{-},0\rangle$ as described for the case of qubit decay. It is worthwhile to note that the mechanism works for any arbitrary initial state including maximally mixed states of the qubits.
\par
%
%
%
%
\par
\emph{Steady state properties and optimization}:
%
\begin{figure*}[t!]
\centering
\includegraphics[width=0.96\textwidth]{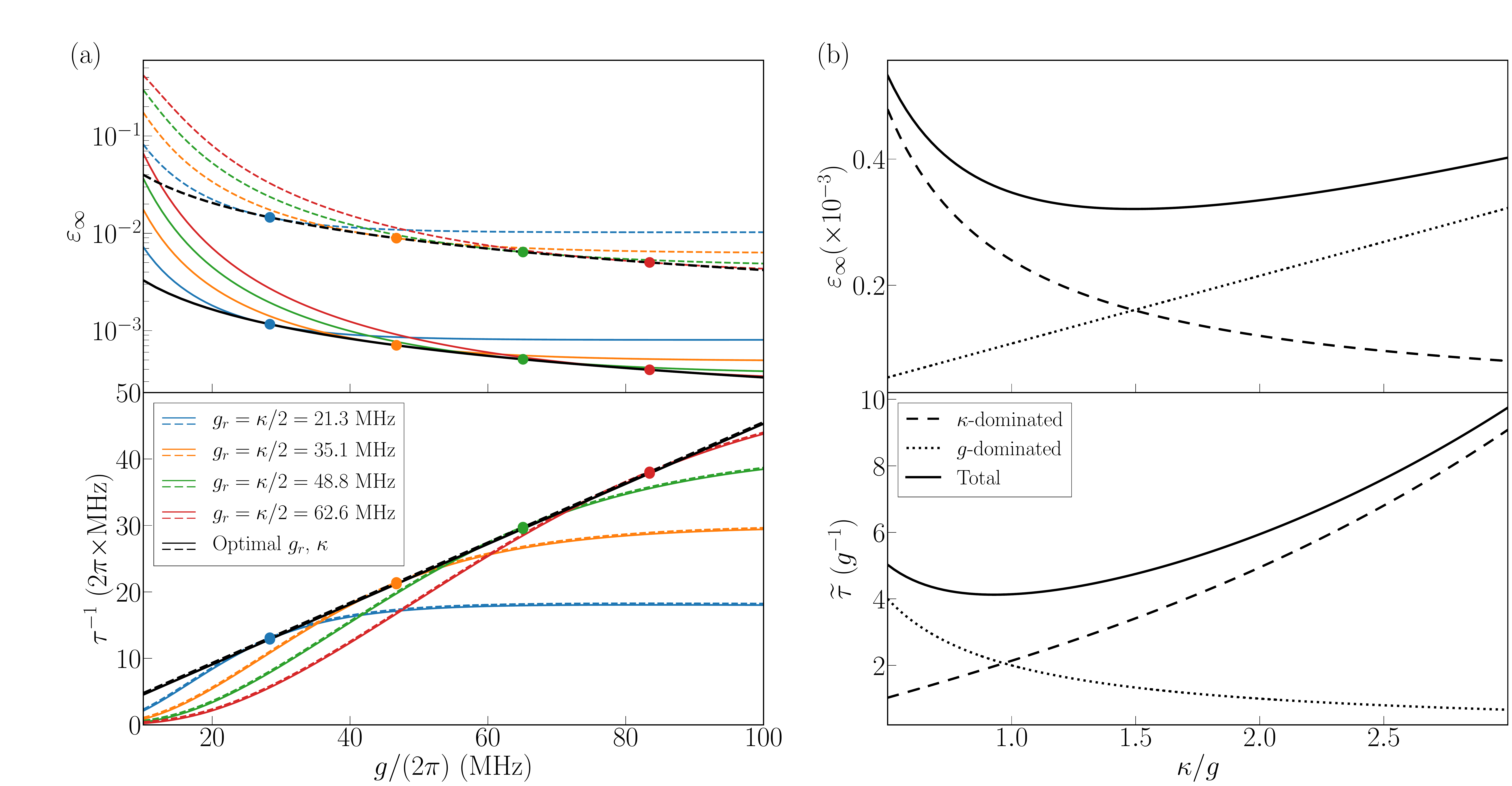}
\caption{(a) Numerical estimates of the steady-state error and convergence rate for the scheme depicted in Fig. \ref{Fig:Mechanism}, calculated as a function of the parametric coupling strength $g$. The black lines show the results for optimal parameters, $\kappa =2 g_{r} = (3/2) g$, with each of the colored dots indicating the value of $g$ for which $(g_r,\kappa)$ are optimal. The solid lines are calculated for $T_{1}= 100\;\mu {\rm s}, T_{2} = 200\;\mu {\rm s}$ (``best case"), while the dashed lines are calculated for  $T_{1}= 10\;\mu {\rm s}, T_{2} = 10\;\mu {\rm s}$ (``worst case"). The qubit decoherence rates affect the target fidelity, but not the convergence time to target state. (b) Analytical estimates of the total steady-state error, denoted by  $\varepsilon_\infty$, and lifetime of the state $|C,0\rangle$ decaying into $|\xi,0\rangle$, denoted by $\widetilde{\tau}$. Each plot also shows the result calculated in the limit $\kappa \gg g$ ($\kappa$-dominated) and in the limit $\kappa \ll g$ ($g$-dominated), for a fixed ratio $g_{r}/g = 3/4$ \cite{Supplement}.}
\label{Fig:Performance}
\end{figure*}
The steady-state dynamics of the full qubits-resonator system can be described using a Liouvillian $\dot{\rho}(t) = \mathcal{L}\rho(t)$, where
\begin{eqnarray}
\mathcal{L} &=& -i [\mathcal{H}_{\rm eff}, \bullet] + \frac{\kappa}{2} \mathcal{D}[a]\bullet \nonumber\\
&& \quad \quad + \sum_{j=1,2} \left(\frac{\gamma_{1}^{j}}{2} \mathcal{D}[\sigma_j]\bullet + \frac{\gamma_{\phi}^{j}}{2} \mathcal{D}[Z_j]\bullet \right) 
\label{eq:ME}
\end{eqnarray}
with $\mathcal{D}[o] \bullet = 2o\bullet o^{\dagger} - \{o^{\dagger}o,\bullet\}$ being the usual Lindblad superoperators describing Markovian decay dynamics for the two qubits and the resonator. Here $\kappa$ denotes the bare resonance linewidth of the resonator, while $(\gamma_{1}^{j}, \gamma_{\phi}^{j})$ denote the qubit relaxation and dephasing rates respectively. We note that use of a local Markovian dissipator in Eq. (\ref{eq:ME}), describing the dominant decay through the resonator acting as an engineered bath for qubits, is motivated by the dispersive circuit-QED regime considered here, $\gamma_{1}^{j}\ll\kappa, g_{jk} \ll \Delta_{jr}$ \cite{Supplement}. The first inequality ensures that the density of states of the resonator appears flat (frequency-independent) over qubit response times, while the second inequality ensures that transitions between widely separated frequencies remain off-resonant in the presence of interaction-induced energy-level splittings. In the presence of colored noise or ultra-strong qubit-resonator coupling \cite{Casanova2010}, the rotating-wave approximation breaks down, leading to unphysical predictions \cite{Levy2014}. In such cases, the master equation needs to be derived in the dressed basis of the qubit-resonator system in order to identify the correct frequency-dependent dissipation rates \cite{Beaudoin2011}.

%
%
\par
Figure \ref{Fig:Performance}(a) shows the results obtained using a full Liouvillian-based numerical optimization \cite{Supplement}. Here, for simplicity of presentation, we have assumed symmetric qubit-qubit couplings $g_{12}^{\pm} = g$ and qubit-resonator couplings $g_{jr}^{\pm} = g_r$. A unique feature of our scheme is the simultaneous reduction in steady state error, $\varepsilon_\infty$, and preparation time, $\tau$, scaling as $1/g$ when $\kappa = 2g_{r} = (3/2)g$. 
%
%
To understand this further, we perform an analytical calculation of steady state error and stabilization time for $|\Phi_{-}\rangle$ (starting from $|eg\rangle$), using a Liouvillian truncated at lowest two resonator levels. The solid-black curves in Fig. \ref{Fig:Performance}(b) present the results of this calculation, and verify that our scheme enables a simultaneous reduction of $\varepsilon_\infty$ and $\tau$ such that 
%
\begin{eqnarray}
     \left(\frac{\varepsilon_\infty}{\tau}\right) = \gamma_{1} F \left[\left(\frac{g_{r}}{g}\right),\left(\frac{\kappa}{g}\right)\right].
\end{eqnarray}
%
For the optimal ratio of $\kappa/g$ (with $\kappa = 2 g_{r}$), indicated with colored dots for different values of $\kappa$ in Fig. \ref{Fig:Performance}(a), dissipation-induced errors and coupling-induced errors can be balanced to attain the minimum steady-state error; here both the steady state error and the preparation time scale together as $1/g$ \cite{Supplement}. This is in contrast to typical stabilization protocols, which operate in either the $\kappa$-dominated regime \cite{Leghtas2013} (decreasing $\varepsilon_\infty$, increasing $\tau$; dashed-black curves) or the $g$-dominated regime \cite{Reiter2013} (increasing $\varepsilon_\infty$, decreasing $\tau$; dotted-black curves), and are constrained to maintain the product $\varepsilon_\infty \tau$ constant. 
\par
Numerical studies also show that the metrics obtained above are robust to imperfections. High-fidelities in excess of 99\% are achievable within few 100 ns, for as large as 50\% deviations from optimal values of $g_{r}$ or $\kappa$, or for asymmetries in the qubit-qubit couplings $g_{12}^{+} \neq g_{12}^{-}$. The performance of the scheme is, however, more sensitive to asymmetries in the parametric qubit-resonator couplings $g_{1r} \neq g_{2r}$ as this introduces a spurious coherent coupling out of the target state. Such spurious couplings may also arise due to imbalance in qubit detunings; specifically, maximum steady-state fidelities are achieved for odd-parity states when $\delta_1 = \delta_2$, and for even-parity states when $\delta_1 =- \delta_2$ \cite{Supplement}. Other sources of error are the counter-rotating terms, which lead to leakage out of $|\xi,0\rangle$ and compromise the parallel scaling of error and stabilization time. Detailed numerical simulations including the dominant leakage channel in the zero-photon manifold show that the effect of such terms can be suppressed and fidelity can be recovered, either by employing sufficiently large dispersive shifts \cite{Supplement} or by the application of optimal control techniques \cite{Reiter2018,FutureWork}.
%
\par
\emph{Continuous-wave Coherent Control}: 
%
%
\begin{figure}[t!]
\centering
\includegraphics[width=\columnwidth]{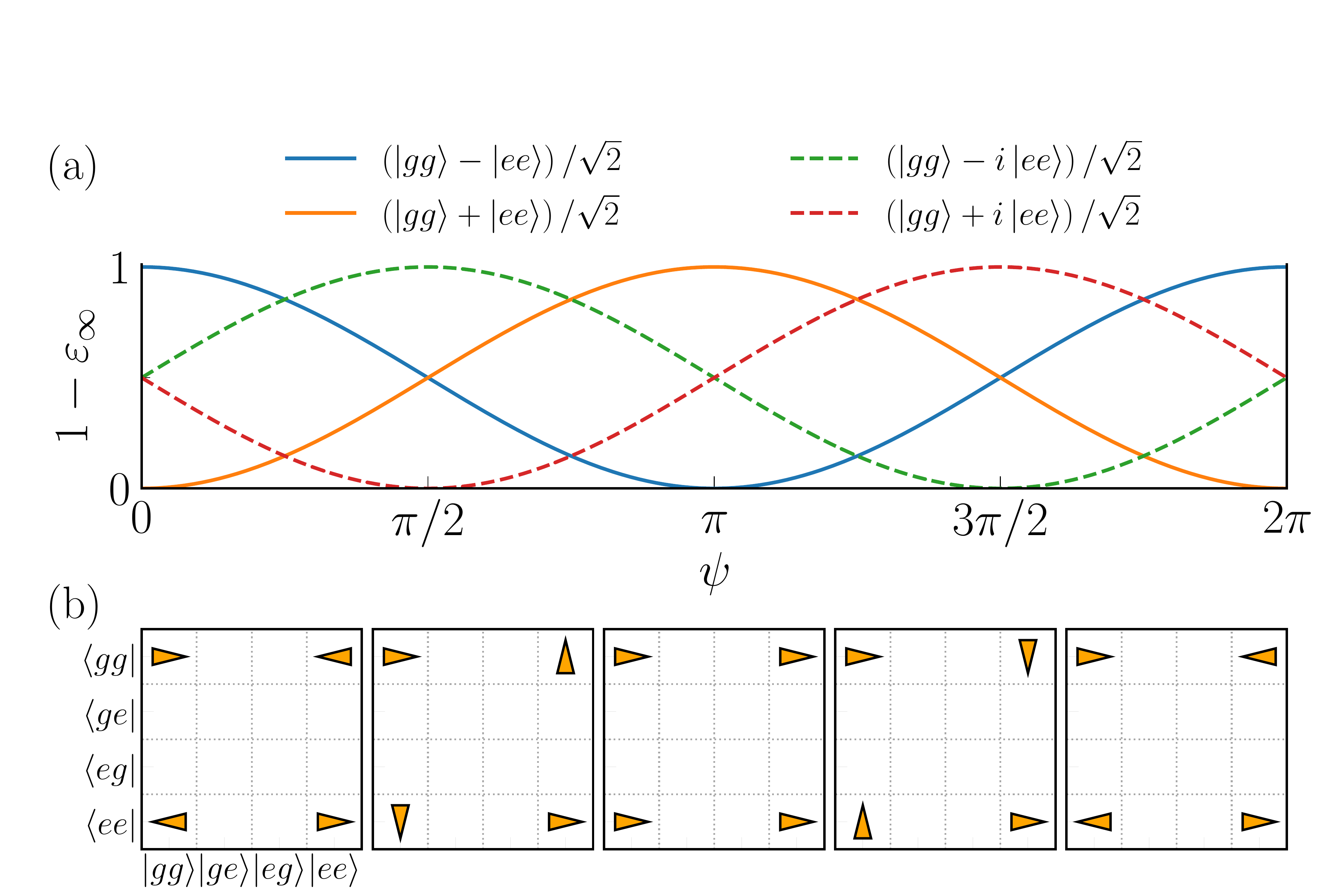}
\caption{Continuous rotations in the stabilized manifold of maximally-entangled even-parity states. (a) Steady state fidelities calculated from the Liouvillian, $\mathcal{L}\rho_{ss}=0$, as a function of $\psi = \phi_{2r} = \pi/2 - \phi_{12}^{+}$. (b) Tomograms of $\rho_{ss}$ for $\psi = k\pi/2$. The opacity and direction of the arrows indicates the magnitude and phase of the matrix elements respectively. All calculations were done for ${T_{1}=100\;\mu {\rm s}},\;{T_{2}=200\;\mu {\rm s}},\;{g = 2\pi\times50\;{\rm MHz}}$ and with optimal values for the other parameters $\kappa^{\rm opt} = 2g_r^{\rm opt} = (3/2)g$.}%
\label{Fig:Gate}%
\end{figure}
In addition to their strength being tunable with pump amplitudes, the phases of the parametric couplings are determined by the phases of the pumps mediating the respective interactions. This provides a unique knob to implement rotations in the two-qubit stabilized subspace. For instance, Table \ref{Table:Bell} shows that the phase entering the collapse operator can be tuned continuously with the phases of the pumps mediating the qubit-resonator couplings. Figure \ref{Fig:Gate} demonstrates how this can be exploited for continuous control of maximally-entangled even-parity states by tuning qubit-qubit coupling phases in tandem with qubit-resonator coupling phases. Notably, this rotation of the target state within a given parity manifold maintains the total population and the purity in the manifold constant. 
%
\par
\emph{Circuit implementation}:
%
%
\begin{figure}[t!]
\centering
\includegraphics[width=\columnwidth]{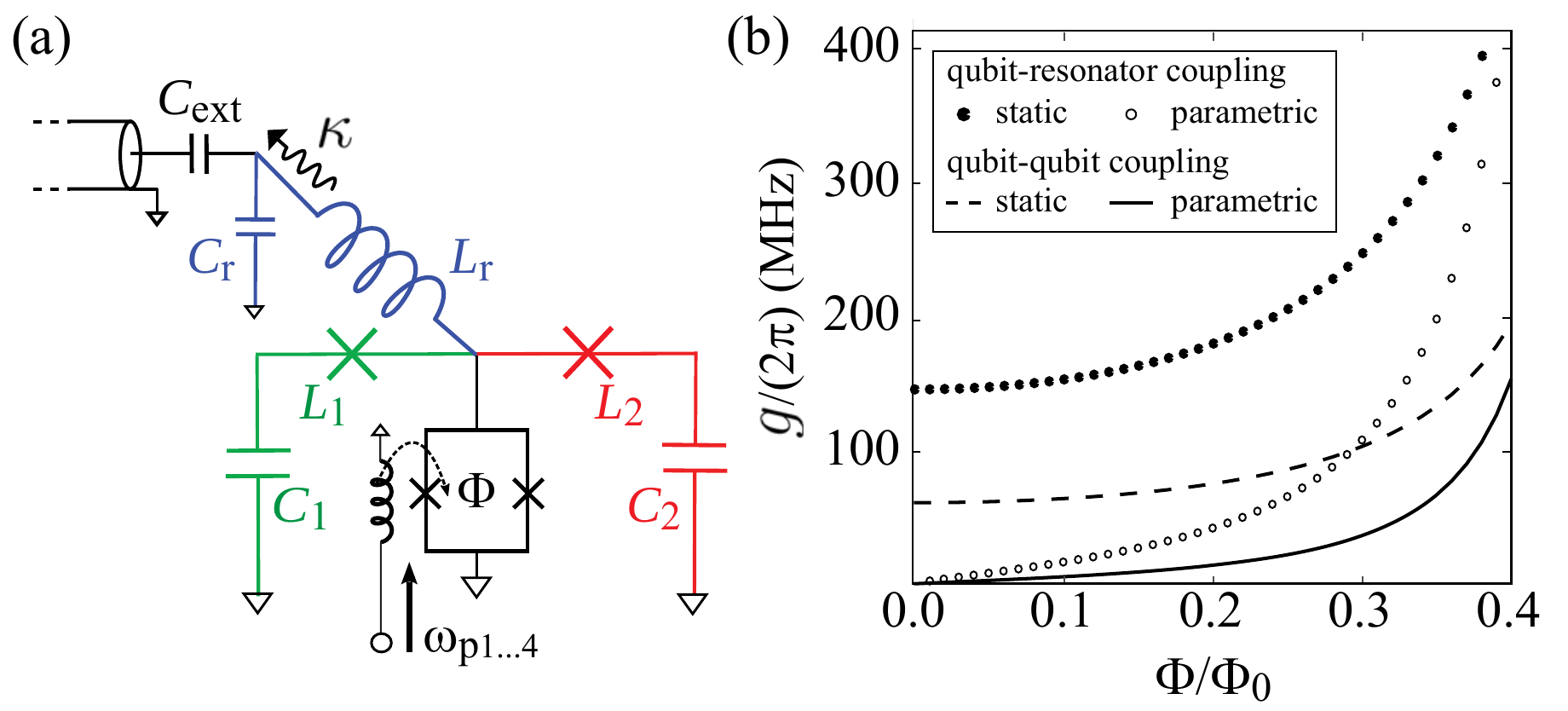}
\caption{(a) Circuit implementation of the 2-qubit Bell state stabilization protocol consisting of two superconducting qubits and a microwave resonator coupled in a T-configuration to a dc-SQUID, driven via an external flux line. (b) Simulation results for qubit-resonator and qubit-qubit static and parametric coupling rates, as a function of the flux through the SQUID loop. The simulation parameters were $E_{J1}/E_{C1}=20, E_{J2}/E_{C2}=30$, $L_r$=5~nH, ${\omega_{1}=2\pi\times4\;{\rm GHz}}$, ${\omega_{2}=2\pi\times6\;{\rm GHz}}$, ${\omega_{r}=2\pi\times10\;{\rm GHz}}$, and $\delta\Phi = 0.1\Phi_0$.}%
\label{Fig:Coupler}%
\end{figure}
In Fig.~\ref{Fig:Coupler}(a), we present a circuit-QED implementation of the dissipative stabilization scheme presented here. In our proposed implementation, two superconducting qubits and a microwave LC resonator share a SQUID-based tunable inductance. The microwave resonator is coupled to an external superconducting transmission line via a coupling capacitor to provide the required linewidth $\kappa$. The tunable parametric coupling is realized through modulation of the SQUID inductance with an external flux line according to $L_{sq}=\Phi_0/(2\pi I_c\cos(\pi\Phi/\Phi_0))$, where $I_c$ is the critical current of the Josephson junctions, $\Phi$ is the flux threading the squid loop and $\Phi_0$ is the magnetic flux quantum. This leads to a static coupling rate $g_{jr}$ between the qubit~$j$ and the resonator as \cite{zakka2011quantum,Lu2017}
\begin{equation}
g_{jr}(\Phi) \approx 
\frac{L_{sq}(\Phi)}{2\sqrt{L_rL_{j}}}\sqrt{\omega_{j}\omega_r},
\end{equation}
where $L_{j},\omega_{j}$ denote the inductance and plasma frequency of the {$j$-th} transmon, while $L_r,\omega_r$ denote the inductance and frequency of the resonator. We see that the coupling rate is proportional to the participation ratio between the SQUID and qubit/resonator inductance. For small flux modulation amplitudes, the parametric coupling rate resulting from a flux drive $\Phi(t)=\delta\Phi_{jr}\cos(\omega_{pj}t+\phi_{pj})$ is given by $g_{jr}=(\partial g_{jr}/\partial \Phi) \delta \Phi_{jr}$. Figure~\ref{Fig:Coupler}(b) shows a simulation of the qubit-qubit and qubit-resonator coupling rates as a function of flux, for a typical set of circuit-QED parameters. The static coupling rates can be easily tuned over more than 100 MHz, for a flux modulation within half-a-period of flux quantum. In the coupler design presented here, $L_{sq}\ll L_{j}$ to protect the qubits against flux noise from the SQUID \cite{Supplement}. Further in this architecture, the self-resonance frequency of the coupler can be designed to be much higher ($\sim$ 20-40 GHz) than qubit and resonator frequencies of interest, for typical specific capacitances associated with Aluminum films. 
%
\par
\emph{Discussion}: 
In summary, we have a proposed a scheme that implements fast and robust high-fidelity entanglement preparation using both unitary (qubit-qubit) and dissipative (qubit-resonator) parametric interactions. In particular, our scheme allows simultaneous optimization of fidelity and stabilization rate by balancing the dissipation-induced errors and the coupling-induced errors. This property is rooted in the fact the target state is a dark state of the engineered collapse operator \emph{and} the system Hamiltonian; Thus, there is no drive-dependent loss mechanism out of the target state that competes with the stabilization mechanism. This is in stark contrast to the usual dissipative state preparation protocols, where parasitic couplings out of the target state impose a hierarchy of coupling rates, and low error can be achieved only by driving the system slower than the effective linewidth of the dressed states to maintain resonant driving \cite{ReiterNJP2012,Aron2016}. Such considerations become increasingly crucial as dissipative engineering protocols are extended to distributed quantum systems, where the speed of state transfer needs to beat the decoherence rate due to correlated noise \cite{Zwick2014}. Parametrically-engineered dissipation also opens possibilities for dissipation-mediated quantum control in the stabilized state space, by exploiting the phase-tunability of the target state. 
\par
Additionally, our scheme offers distinct operational advantages over the previously considered schemes for entanglement stabilization in circuit-QED-like platforms. Since there is no direct driving of the bath resonator, the size of excitations in the resonator, $\bar{n} = \langle a^{\dagger} a \rangle$, remains small. This is a favorable situation from the point of view of avoiding measurement-induced dephasing of the qubits due to photon number fluctuations, which grows as $\Gamma_{m}~\sim~\bar{n}\kappa$ \cite{Gambetta2006}. Further, parametric qubit-resonator interactions have no number dependence; hence our scheme employs no photon-selectivity for shuttling excitations across the resonator ladder unlike the usual dispersively coupled schemes \cite{Leghtas2013}. Driving number-selective transitions usually places stringent requirements on matching the dispersive shifts, $|\Delta \chi| \ll \chi^{2}/\kappa \sqrt{\bar{n}}$, which ceases to be a constraint for parametrically-driven qubit-resonator transitions. Finally, since our scheme has no direct qubit driving, there are no associated Stark shifts and thus no power-dependent pump detuning is necessary.
%
\begin{acknowledgments}
We thank Christian Kraglund Andersen, Luke Govia and Lorenza Viola for helpful discussions. This research was supported by the Department of Energy under grant DE-SC0019461. FR acknowledges financial support by a Feodor-Lynen fellowship from the Alexander von Humboldt Foundation.
\end{acknowledgments}
%
%
%
%
%
\end{document}


\title{Supplemental material for\\``High-fidelity dissipative engineering using parametric interactions"}
\author{E. Doucet}
\affiliation{Department of Physics and Applied Physics, University of Massachusetts, Lowell, MA 01854, USA}
\author{F. Reiter}
\affiliation{Harvard University, Department of Physics, 17 Oxford St, Cambridge, MA 02138, USA}
\author{L. Ranzani}
\affiliation{Raytheon BBN Technologies, Cambridge, Massachusetts 02138, USA}
\author{A. Kamal}
\affiliation{Department of Physics and Applied Physics, University of Massachusetts, Lowell, MA 01854, USA}
\affiliation{Research Laboratory of Electronics, Massachusetts Institute of Technology, Cambridge, MA 02139, USA}
\maketitle
\parskip=5 pt
\section{Derivation of Effective Hamiltonian}
\label{sec:HamilDerivation}
%
The main text introduces the following Hamiltonian in the lab frame for parametrically-engineered stabilization 
\begin{equation}
    \mathcal{H} = 
        \frac{\omega_1}{2} Z_1 
      + \frac{\omega_2}{2} Z_2 
      + \omega_r a^\dagger a
      + \tilde{g}_{12}(t) X_1 X_2
      + \tilde{g}_{1r}(t) X_1 X_r
      + \tilde{g}_{2r}(t) X_2 X_r
      ,
\end{equation}
with the time-dependent couplings
\begin{subequations}
    \begin{align}
        \tilde{g}_{12}(t) &\equiv g_{12}^{+} \expss{i \omega_{12}^{+} t} + g_{12}^{-} \expss{i \omega^{-}_{12} t} + \mathit{h.c.},
        \\
        \tilde{g}_{1r}(t) &\equiv g_{1r} \left(\expss{i \omega_{1r}^{+} t} + \expss{i \omega_{1r}^{-} t}\right) + \overline{g}_{1r} + \mathit{h.c.},
        \\
        \tilde{g}_{2r}(t) &\equiv g_{2r} \left(\expss{i \omega_{2r}^{+} t} + \expss{i \omega_{2r}^{-} t}\right) + \overline{g}_{2r} + \mathit{h.c.}
    \end{align}
\end{subequations}
The $\overline{g}_{jr}$ terms correspond to static/dispersive couplings and the remaining terms correspond to time-dependent parametric drives. 
\par
To construct the effective Hamiltonians introduced in the main text as Eqs. (1) and (2), we first perform a Schrieffer-Wolff transformation on the dispersive part of the Hamiltonian to diagonalize the dispersive coupling between the qubits and resonator. Using the convention $Z \ket{g} = +\ket{g}$ and $Z \ket{e} = -\ket{e}$, the Hamiltonian becomes
\begin{align}
	\begin{split}
		\label{eqn:hpostsw}
		\widetilde{\mathcal{H}}
		&= \frac{\omega_1 + 2\chi_1 a^\dagger a}{2} Z_1
	    + \frac{\omega_2 + 2\chi_2 a^\dagger a}{2} Z_2
	    + \omega_r a^\dagger a 
	    + \left(
	       	\tilde{g}_{12}(t) - \frac{\overline{g}_{1r}\overline{g}_{2r}}{\Delta_{1r}} - \frac{\overline{g}_{1r}\overline{g}_{2r}}{\Delta_{2r}}
	      \right) X_1X_2 \\
		&\quad
		+ g_{1r} \left(\expss{i \left(\omega_{1r}^{+} t + \phi_{1r}\right)} + \expss{i \left(\omega_{1r}^{-} t + \phi_{1r}\right)}\right) X_1 X_r	
		+ g_{2r} \left(\expss{i \left(\omega_{2r}^{+} t + \phi_{2r}\right)} + \expss{i \left(\omega_{2r}^{-} t + \phi_{2r}\right)}\right) X_2 X_r
		, \\
	\end{split}
\end{align}
with dispersive shifts $\chi_{j} \equiv \overline{g}_{jr}^2/\Delta_{jr}$ where $\Delta_{jr} \equiv \omega_{r} - \omega_j$. We then move into the interaction frame defined by $
\widetilde{\mathcal{H}}_0 = 
	\left[\left(\omega_1 / 2 \right) + \chi_1 a^\dagger a \right] Z_1 
  + \left[\left(\omega_2 / 2 \right) + \chi_2 a^\dagger a \right] Z_2 
  + \omega_r a^\dagger a
$
via the unitary operator $U = U_f U_1 U_2$ with the commuting factors
\begin{subequations}
	\begin{align}
		U_f &= \exp\left[i \left(\frac{\omega_1}{2} Z_1 + \frac{\omega_2}{2} Z_2 + \omega_r a^\dagger a \right) t \right], \\	
		U_1 &= \exp\left[i \chi_1 a^\dagger a Z_1 t\right], \\	
		U_2 &= \exp\left[i \chi_2 a^\dagger a Z_2 t\right]	
		.
	\end{align}	
\end{subequations}	
We transform each of the three parametric coupling terms in Eq. \eqref{eqn:hpostsw} individually, starting with the qubit-qubit coupling where we find
\begin{align}
	\label{eqn:sigmaxxtransformfree}
	\begin{split}
		U X_1X_2 U^\dagger 
			&= \expss{-i\left( \omega_1 + \omega_2 \right)t} U_1 \sigma_1 U_1^\dagger U_2 \sigma_2 U_2^\dagger
			 + \expss{-i\left( \omega_1 - \omega_2 \right)t} U_1 \sigma_1 U_1^\dagger U_2 \sigma_2^\dagger U_2^\dagger
			 + \textit{h.c.}	
			 ,		 
	\end{split}
\end{align}
since $U_f \sigma_j U_f^\dagger = \exp\left(-i \omega_j t\right)\sigma_j$ and $\comm{U_1}{\sigma_2} = \comm{U_2}{\sigma_1} = 0$. To find $U_1 \sigma_1 U_1^\dagger$, we write $\sigma_1$ in Dirac notation
\begin{align*}
	\sigma_1 \to \sigma_1 \otimes \mathbb{I}_2 \otimes \mathbb{I}_c \equiv \sum_{n=0}^{\infty}\sum_{q_2\in\{g,e\}} \ket{g,q_2,n}\bra{e,q_2,n}
	,
\end{align*}
where we have used the basis $\ket{q_1, q_2, n} \equiv \ket{q_1}\otimes\ket{q_2}\otimes\ket{n}$, with $q_{1,2}\in\{g,e\}$ and $n\in\mathbb{N}$. This gives
\begin{equation}
		U_1 \sigma_1 U_1^\dagger 
			= \sum_{n=0}^{\infty}\sum_{q_2\in\{g,e\}} \expss{2i \chi_1 n t} \ket{g,q_2,n}\bra{e,q_2,n}
			.
\end{equation}
Using this result and the similar result for $U_2 \sigma_2 U_2^\dagger$ in \eqref{eqn:sigmaxxtransformfree}, we find
\begin{align}
	\label{eqn:qqtransform}
	\begin{split}
		U X_1X_2 U^\dagger 
			&= \sum_{n=0}^{\infty} \expss{-i\left[ \omega_1 + \omega_2 - 2n\left(\chi_1 + \chi_2\right)\right] t} \ket{g,g,n}\bra{e,e,n} 
			+ \sum_{n=0}^{\infty} \expss{-i\left[ \omega_1 - \omega_2 - 2n\left(\chi_1 - \chi_2\right)\right] t} \ket{g,e,n}\bra{e,g,n} + \textit{h.c.} \\
	\end{split}
\end{align}
We now turn our attention to transforming the qubit-resonator couplings into the interaction frame
\begin{align}
	\label{eqn:sigmacpcdtransformfree}
	U X_1 X_r U^\dagger
		&= \expss{-i\left( \omega_1 + \omega_r \right)t} U_{12} \sigma_1 a U_{12}^\dagger
		 + \expss{-i\left( \omega_1 - \omega_r \right)t} U_{12} \sigma_1 a^\dagger U_{12}^\dagger
		 + \textit{h.c.}
		 ,
\end{align}
where we have introduced $U_{12} = U_1 U_2 = \exp\left[ i\{\left( \chi_1Z_1 + \chi_2Z_2 \right) a^\dagger a\} t\right]$. As before, to find $U_{12} \sigma_1 a U_{12}$ and $U_{12} \sigma_1 a^\dagger U_{12}$ we write the operators in Dirac notation
\begin{align*}
	\sigma_1 a &\to \sigma_1 \otimes \mathbb{I}_2 \otimes a \equiv \sum_{n=0}^{\infty} \sum_{q_2\in\{g,e\}} \sqrt{n+1}\ket{g,q_2,n}\bra{e,q_2,n+1} 
	,
\end{align*}
giving
\begin{equation}
	\label{eqn:transforms1c}
		U_{12} \sigma_1 a U_{12}^\dagger
			= \sum_{n=0}^{\infty} \sum_{q_2\in\{g,e\}} \sqrt{n+1}
				\expss{-i \left[-\chi_1(2n+1) + \chi_2\expval{Z_2}{q_2}\right] t}
				\ket{g,q_2,n}\bra{e,q_2,n+1} 
			.
\end{equation}
Substituting this into \eqref{eqn:sigmacpcdtransformfree} yields
\begin{align}
	\label{eqn:qctransform}
	\begin{split}
		U \sigma_1 \left( a + a^\dagger \right) U^\dagger
			&= \sum_{n=0}^{\infty} \sum_{q_2\in\{g,e\}} \sqrt{n+1} 
				\expss{-i \left[ \omega_1 + \omega_r - \chi_1(2n+1) + \chi_2\expval{Z_2}{q_2}\right] t}
				\ket{g,q_2,n}\bra{e,q_2,n+1} \\
			&\quad + \sum_{n=0}^{\infty} \sum_{q_2\in\{g,e\}} \sqrt{n+1}
				\expss{-i \left[ \omega_1 - \omega_r - \chi_1(2n+1) - \chi_2\expval{Z_2}{q_2}\right] t}
				\ket{g,q_2,n+1}\bra{e,q_2,n}
				.
	\end{split}
\end{align}
The results in Eqs. \eqref{eqn:qqtransform} and \eqref{eqn:qctransform} allow us to express Eq. \eqref{eqn:hpostsw} in the interaction frame
\begin{align}
	\begin{split}
		\widetilde{\mathcal{H}}_I
			&= \left(
		        g_{12}^+ \expss{i \left(\omega_{12}^+ t + \phi_{12}^{+}\right)} + g_{12}^- \expss{i \left(\omega_{12}^- t + \phi_{12}^{-}\right)}
		        - \frac{\overline{g}_{1r}\overline{g}_{2r}}{\Delta_{1r}} - \frac{\overline{g}_{1r}\overline{g}_{2r}}{\Delta_{2r}}
		    	   \right) 
		    	   \left[
		    	    \sum_{n=0}^{\infty} \expss{-i\left[ \omega_1 + \omega_2 - 2n\left(\chi_1 + \chi_2\right)\right] t} \ket{g,g,n}\bra{e,e,n} \right. \\
			    & \hspace{7cm} \left. + 
			    \sum_{n=0}^{\infty} \expss{-i\left[ \omega_1 - \omega_2 - 2n\left(\chi_1 - \chi_2\right)\right] t} \ket{g,e,n}\bra{e,g,n}
			   \right] \\
			&+ g_{1r} \left(\expss{i \left(\omega_{1r}^{+} t + \phi_{1r}\right)} + \expss{i \left(\omega_{1r}^{-} t + \phi_{1r}\right)}\right)
			   \left[
			   	\sum_{n=0}^{\infty} \sum_{q_2\in\{g,e\}} \sqrt{n+1} 
				\expss{-i \left[ \omega_1 + \omega_r - \chi_1(2n+1) + \chi_2\expval{Z_2}{q_2}\right] t}
				\ket{g,q_2,n}\bra{e,q_2,n+1} \right. \\
				& \hspace{4cm} \left. + 
				\sum_{n=0}^{\infty} \sum_{q_2\in\{g,e\}} \sqrt{n+1}
				\expss{-i \left[ \omega_1 - \omega_r - \chi_1(2n+1) - \chi_2\expval{Z_2}{q_2}\right] t}
				\ket{g,q_2,n+1}\bra{e,q_2,n}
			   \right] \\
			&+ g_{1r} \left(\expss{i \left(\omega_{1r}^{+} t + \phi_{1r}\right)} + \expss{i \left(\omega_{1r}^{-} t + \phi_{1r}\right)}\right)
			   \left[
			   	\sum_{n=0}^{\infty} \sum_{q_1\in\{g,e\}} \sqrt{n+1} 
				\expss{-i \left[ \omega_2 + \omega_r - \chi_2(2n+1) + \chi_1\expval{Z_1}{q_1}\right] t}
				\ket{q_1,g,n}\bra{q_1,e,n+1} \right. \\
				& \hspace{4cm} \left. + 
				\sum_{n=0}^{\infty} \sum_{q_1\in\{g,e\}} \sqrt{n+1}
				\expss{-i \left[ \omega_2 - \omega_r - \chi_2(2n+1) - \chi_1\expval{Z_1}{q_1}\right] t}
				\ket{q_1,g,n+1}\bra{q_1,e,n}
			   \right] + \textit{h.c.} \\
	\end{split}
\end{align}
The main text states the frequencies necessary to target any maximally entangled state of a fixed parity. Here we demonstrate the Hamiltonian constructed to stabilize an odd-parity state by choosing
\begin{eqnarray*}
	& &\omega_{12}^+ = \omega_1 + \omega_2; \hspace{7.6em} \omega_{1r}^\pm = \omega_1 + \omega_r - \chi_1 \pm \chi_2 \\ 
	& & \omega_{12}^- = \omega_1 - \omega_2 - 2\chi_1 + 2\chi_2; \qquad \omega_{2r}^\pm = \omega_2 + \omega_r - \chi_2 \pm \chi_1.
\end{eqnarray*}
For this choice of pump frequencies, we obtain
\begin{align}
	\begin{split}
		 \widetilde{\mathcal{H}}_I &= 
			   g_{12}^{+}e^{i\phi_{12}^{+}} \left\{ \ket{g,g,0}\bra{e,e,0} + \expss{-2i\left(\chi_1 + \chi_2 \right)t} \ket{g,g,1}\bra{e,e,1} \right\} \\
			&\quad + g_{12}^{-}e^{i\phi_{12}^{-}} \left\{ \expss{+2i\left(\chi_1 - \chi_2 \right)t} \ket{g,e,0}\bra{e,g,0} + \ket{g,e,1}\bra{e,g,1} + \expss{-2i\left(\chi_1 - \chi_2 \right)t} \ket{g,e,2}\bra{e,g,2}\right\} \\
			&\quad + g_{2r}^{+}e^{i\phi_{2r}^{+}} \left\{ \ket{g,g,0}\bra{g,e,1} + \ket{e,g,0}\bra{e,e,1} + \expss{+2i \chi_1 t} \ket{g,g,0}\bra{g,e,1} + \expss{-2i \chi_1 t} \ket{e,g,0}\bra{e,e,1} \right\} \\
			&\quad + g_{1r}^{+}e^{i\phi_{1r}^{+}} \left\{ \ket{g,g,0}\bra{e,g,1} + \ket{g,e,0}\bra{e,e,1} + \expss{+2i \chi_2 t} \ket{g,g,0}\bra{e,g,1} + \expss{-2i \chi_2 t} \ket{g,e,0}\bra{e,e,1} \right\} + \textit{h.c.} 
			,
			\\
	\end{split}
	\label{eqn:HNoRwa}
\end{align}
where we have kept only the static and the dominant counter-rotating terms. The static part of this Hamiltonian, with the resonator truncated to its lowest two energy levels, gives the effective Hamiltonian reported in the main text as Eq. (2)
\begin{equation}
    \mathcal{H}_{\rm eff}^{\rm odd} = g_{12}^{+}e^{i\phi_{12}^{+}}(1 - \langle a^{\dagger} a \rangle)\sigma_{1} \sigma_{2} + g_{12}^{-}e^{i\phi_{12}^{-}}\langle a^{\dagger} a \rangle \sigma_{1} \sigma_{2}^{\dagger} + (g_{1r}^{+}e^{i\phi_{1r}^{+}}\sigma_{1}^{\dagger} + g_{2r}^{+}e^{i\phi_{2r}^{+}}\sigma_{2}^{\dagger}) a^{\dagger} + h.c.,
\end{equation}
where we have again expressed the Hamiltonian in terms of canonical operators in this truncated space.
%
\section{Justification for Master Equation}
%
In this section we outline the justification behind using the master equation used in the main text. Since $\kappa \gg \gamma_{1,2}$, the analysis presented below is to quantify the effect of the dominant dissipation channel (i.e. resonator decay) on the system of two qubits in the presence of strong parametric coupling between the qubits and the resonator. To this end, considering a bosonic environment for the resonator, we can write
\begin{eqnarray}
& & \mathcal{H}_{S} = 
        \frac{\omega_1}{2} Z_1 
      + \frac{\omega_2}{2} Z_2 
      + \omega_r a^\dagger a
      + \tilde{g}_{12}(t) X_1 X_2
      + \tilde{g}_{1r}(t) X_1 X_r
      + \tilde{g}_{2r}(t) X_2 X_r,\\
& & \mathcal{H}_{E} = \sum_{\alpha} \nu_{\alpha} r_{\alpha}^{\dagger} r_{\alpha},\\
& & \mathcal{H}_{SE} =\sum_{\alpha} \mu_{\alpha}^{r} r_{\alpha}^{\dagger} a + h.c.
\end{eqnarray}
Following the same steps as followed for transforming $\mathcal{H}_{S}$ in Sec. \ref{sec:HamilDerivation}, we first diagonalize the static dispersive coupling by performing Schrieffer-Wolff Transformation. To leading order in the parameter $g_{jr}/\Delta_{jr}$, this transforms the system -environment Hamiltonian as
\begin{eqnarray}
\widetilde{\mathcal{H}}_{SE} = \sum_{\alpha} \mu_{\alpha}^{r} r_{\alpha}^{\dagger} \left(a - \sum_{j =1,2} \frac{\bar{g}_{jr}(t)}{\Delta_{jr}} \sigma_{j}\right) + h.c.
\end{eqnarray}
Next, moving into a rotating frame for both the system and environment this gives,
\begin{eqnarray}
\widetilde{\mathcal{H}}_{SE, I} = \sum_{\alpha}  \mu_{\alpha}^{r} r_{\alpha}^{\dagger} a  e^{i(\nu_{\alpha} -\omega_{r}-2\sum_{j =1,2}\chi_{j}\langle\sigma_{j}^{Z}\rangle)t} -  \sum_{\alpha} \sum_{j =1,2} \mu_{\alpha}^{r} \frac{\bar{g}_{jr}}{\Delta_{jr}}  r_{\alpha}^{\dagger}  \sigma_{j} e^{i(\nu_{\alpha} -\omega_{j} - 2 \langle a^{\dagger} a \rangle \chi_{j})t}.
\end{eqnarray}
with qubit frequencies incorporating the corresponding static dispersive shifts alone, $\chi_{j} = \bar{g}_{jr}^{2}/\Delta_{jr}$. Reintroducing the parametric interaction as a time-dependent perturbation on the mixing coefficient, the above interaction can be rewritten in the form
\begin{eqnarray}
\widetilde{\mathcal{H}}_{SE,I} = R(t) \Gamma^{\dagger}(t) + \sum_{j =1,2} Q_{j}(t) \Gamma^{\dagger}(t) + h.c.
\label{Eq:HSEint}
\end{eqnarray}
where
\begin{eqnarray*}
& & R(t) = a e^{-i \omega_{r}'t}, \quad Q(t) = \left(\frac{\tilde{g}_{jr}(t)}{\Delta_{jr}}\right) \sigma_{j}  e^{-i\omega_{j}'t}; \quad \Gamma(t) = \sum_{\alpha} \mu_{\alpha}^{r*} r_{\alpha} e^{-i\nu_{\alpha}t}.
\end{eqnarray*}
with $\omega_{r}' = \omega_{r}+2\sum_{j =1,2}\langle\sigma_{j}^{Z}\rangle \chi_{j}$ and $\omega_{j}' = \omega_{j} + 2\langle a^{\dagger} a \rangle\chi_{j}$. Note that we have ignored the inertial term that arises if one performs a full time-dependent Schrieffer-Wolff transformation which leads to additional dispersive shifts due to parametric interactions between the qubits and resonator; this approximation is valid in the RWA regime \cite{ZhihaoTDSWT}, $|g_{jr}| \ll \chi < \bar{g}_{jr} \ll \Delta_{jr}$ (white region in Fig.~\ref{Fig:Scaling}), required to maintain the concurrent scaling of target fidelity and stabilization rate.  
\par
Assuming Markovian evolution and taking the zero temperature limit for the environment \cite{CarmichaelVol1}, leads to the following terms in the master equation,
\begin{eqnarray}
\dot{\rho}_{I} (t) & = &\left(a \rho_{I} (t) a^{\dagger}- a^{\dagger} a\rho_{I} (t)\right)\int_{0}^{t} dt' e^{-i\omega_{r}'(t'-t)}\langle \Gamma(t)\Gamma^{\dagger}(t')\rangle \nonumber\\
& & + \sum_{j =1,2} \left(\frac{\bar{g}_{jr}}{\Delta_{jr}}\right)^{2}  \left(\sigma_{j}\rho_{I} (t) \sigma_{j}^{\dagger} - \sigma_{j}^{\dagger} \sigma_{j}\rho_{I} (t)\right) \int_{0}^{t} dt'e^{-i\omega_{j}'(t'-t)}\langle \Gamma(t)\Gamma^{\dagger}(t')\rangle \nonumber\\
& &+ \sum_{j =1,2} \left|\frac{g_{jr}}{\Delta_{jr}}\right|^{2} \left(\sigma_{j} \rho_{I} (t) \sigma_{j}^{\dagger} - \sigma_{j}^{\dagger}\sigma_{j}\rho_{I} (t)\right)\int_{0}^{t} dt' \left(e^{-i(\omega_{j}' + \omega_{jr})(t'-t)}+e^{-i(\omega_{j}' - \omega_{jr})(t'-t)}\right)\langle \Gamma(t)\Gamma^{\dagger}(t')\rangle \nonumber\\
& & +\; h.c.,
\label{Eq:MEint}
\end{eqnarray}
Here we have decomposed the qubit-resonator interaction into two parts, one corresponding to the static qubit-resonator coupling and the other due to the time-varying parametric part $g_{jr}(t) = g_{jr} e^{i(\omega_{jr}t + \phi_{jr})} + h.c.$. Introducing a continuous spectral density of states for the environment,
\begin{eqnarray*}
\langle \Gamma(t)\Gamma^{\dagger}(t')\rangle  = \int_{0}^{\infty} d \nu \, e^{-i\nu (t-t')} \rho(\nu) |\kappa (\nu)|^{2},
\end{eqnarray*}
and using it in Eq. (\ref{Eq:MEint}) we obtain
\begin{eqnarray}
\dot{\rho}_{I} (t) & = & - i [\Delta, \rho_{I}(t)] + \kappa \mathcal{D}[a]\rho_{I}(t) + \sum_{j =1,2} \kappa \left(\frac{\bar{g}_{jr}}{\Delta_{jr}}\right)^{2} \mathcal{D}[\sigma_{j}]\rho_{I}(t)  + \sum_{j =1,2}  \kappa'\left|\frac{g_{jr}}{\Delta_{jr}}\right|^{2}  \mathcal{D}[\sigma_{j}]\rho_{I}(t). 
\label{Eq:MEintfinal}
\end{eqnarray}
Here we have used
\begin{eqnarray*}
\int_{0}^{t} dt'  e^{-i(\omega-\nu)(t-t')} = \pi \delta (\omega - \nu) + i P\left(\frac{1}{\omega - \nu}\right)
\end{eqnarray*}
with the corresponding decay rates and lamb shift (usually absorbed by redefining bare resonance frequencies) defined as
\begin{eqnarray*}
	& & \kappa(\omega) = 2\pi \rho(\omega)|\mu^{r}(\omega)|^{2}; \qquad \kappa'(\omega) = 2\pi \rho(\omega \pm \omega_{p})|\mu^{r}(\omega\pm \omega_{p})|^{2}; \qquad \Delta = \frac{P}{2\pi} \int_{0}^{\infty} d\nu \frac{\kappa(\omega)}{\omega - \nu}.
\end{eqnarray*}
If the decay is calculated for the resonator the corresponding density of states needs to be calculated at the resonator frequency $\omega_{r}'$, while for the induced decay on the qubit the spectral density of the environment near $\omega_{j}'$ is relevant. As is evident from Eq. (\ref{Eq:MEintfinal}), the parametric coupling contributes towards qubit decoherence with a term almost identical to that due to the dispersive-coupling induced Purcell decay, except that the relevant environmental modes that contribute are shifted by $\pm \omega_{p}$. Depending on the choice of pump frequencies, only one of the up/down-converted noise terms maybe relevant; for instance, for the qubit-resonator pump frequencies stated in Sec.~\ref{sec:HamilDerivation}, only the $\omega_{j}' + \omega_{jr}$ contributes towards parametrically-induced spontaneous emission. Further, under the assumption that the environment spectrum is white, $\kappa \approx \kappa'$, leading to a simple modifiation of the Purcell decay now including the contribution due to parametric coupling.
\par
Doing a similar analysis for the qubit-environment coupling under Schrieffer-Wolff transformation, 
\begin{eqnarray}
\widetilde{\mathcal{H}}_{SE}=  \sum_{j =1,2}\sum_{\alpha} \mu_{\alpha,j}^{q} q_{\alpha,j}^{\dagger} \left(\sigma_{j} + \frac{\bar{g}_{jr}(t)}{\Delta_{jr}} a Z_{j}\right) + h.c.
\label{Eq:HSEintqubit}
\end{eqnarray}
leads to the following damping superoperators 
\begin{eqnarray}
\sum_{j =1,2}\gamma \mathcal{D}[\sigma_{j}]\bullet+ \gamma \left(\frac{\bar{g}_{jr}}{\Delta_{jr}}\right)^{2} \mathcal{D}[aZ_{j}]\bullet + \gamma'\left|\frac{g_{jr}}{\Delta_{jr}}\right|^{2}  \mathcal{D}[aZ_{j}]\bullet.
\label{Eq:MEintfinal2}
\end{eqnarray}
As before, the additional dissipators are weighted by a factor proportional to $|g_{jr}/\Delta| \ll1$ and the primary difference due to parametric coupling is the frequency at which environmental noise is sampled: while $\gamma$ corresponds to noise at respective resonant frequencies, $\gamma'$ is determined by noise at the qubit frequency shifted by $\pm \omega_{p}$. 
\par
%
\emph{Heating}:
%
Usually terms of type $r_{\alpha}^{\dagger} a^{\dagger}, \; q_{\alpha,j}^{\dagger} \sigma_{j}^{\dagger}$ in the system-environment interaction correspond to negative frequency terms in the environment spectrum, which have zero spectral weight and hence do not contribute towards spontaneous emission. However, in the presence of parametric driving, the environmental modes down to $-\omega_{jr}$ are accessible, leading to environment-mediated excitations of the qubit and resonator even at zero temperature. These processes lead to superoperators of the form
\begin{eqnarray}
 	\sum_{j =1,2}  \kappa'\left|\frac{g_{jr}}{\Delta_{jr}}\right|^{2}  \mathcal{D}[\sigma_{j}^{\dagger}]\bullet, \quad  	\sum_{j =1,2}  \gamma'\left|\frac{g_{jr}}{\Delta_{jr}}\right|^{2}  \mathcal{D}[a^{\dagger}]\bullet.
\end{eqnarray}
Nonetheless, in the dispersive regime such environment-induced heating of the system would remain negligible and is hence neglected in the analysis presented in the main text.
%
\section{Analytical calculation and parameter optimization}
%
In this section we present an analytical estimation of the steady state error, extracted from the steady state density matrix $\rho_{ss}$ obtained from the Liouvillian as $\mathcal{L}\rho_{ss} = 0$. By truncating the resonator Hilbert space to include two lowest resonator levels, we obtain a closed set of equations, which can be further constrained by assuming that mixing of different states (off-diagonal elements of $\rho_{ss}$) due to qubit decays can be ignored in the regime of interest i.e. $\gamma \ll (g, g_{r}, \kappa)$. Imposing a relation between qubit-qubit and qubit-resonator couplings, such as $g_r = (3/4)g$, and performing an expansion to leading order in the relevant small parameter, we obtain the following expressions for steady state error in the $\kappa$-dominated (or strong dissipation) and $g$-dominated (or weak dissipation) regimes, respectively:
\begin{eqnarray}
\varepsilon_{\infty} = \left\{
                            \begin{array}{cc}
                            12.6 \displaystyle\left(\frac{\gamma_{1}}{\kappa}\right);& \kappa \gg g\\
                            \\
                            22.4 \displaystyle\left(\frac{1}{\mathcal{C}}\right);& \kappa \ll g,
                            \end{array}
                        \right.
  \label{Eq:AnalyticalError}  
\end{eqnarray}
where $\mathcal{C} = (4g^{2}/\kappa \gamma_{1})$ is identified as the \emph{cooperativity} of the system. Minimizing the total error as the sum of two contributions leads to an optimality condition for the ratio $\kappa/g = (3/2)$ with the minimum error that scales inversely with parametric coupling strength $g$
\begin{eqnarray}
    \varepsilon_{\infty}^{\rm min} \approx 16.8 \displaystyle\left(\frac{\gamma_{1}}{g}\right).
      \label{Eq:AnalyticalErrorOpt}  
\end{eqnarray}
The optimal ratio between dissipative and coherent couplings is in excellent agreement with the value obtained using a full numerical optimization with the given choice of coupling ratio.
\par
The calculation of convergence time $\tau$ is more complicated since it is derived from the relevant spectral gap which involves inverting a typically large matrix. For instance, even for a rather conservative system comprising two qubits and a resonator truncated at two levels, an exact calculation involves diagonalization of a $64 \times 64$ Liouvillian. Nonetheless, an analytical estimate and qualitative scaling with parameters can be obtained by identifying a minimal decoupled subspace including the target state, and self-consistently solving for the populations in different states spanning the subspace. The states $\{|C,0\rangle, |\bar{\xi},1\rangle, |\xi,1\rangle\}, |\xi,0\rangle\}$ form such a subspace (Fig. 1 of the main paper), in which we can write the rate of preparation of the target state $|\xi,0\rangle$ as
\begin{eqnarray}
    \dot{P}_{|\xi,0\rangle}(t) = \Gamma_{C,0} P_{|C,0\rangle} 
                            + \Gamma_{\bar{\xi},1}  P_{ |\bar{\xi},1\rangle}
                            + \Gamma_{\xi,1}  P_{ |\xi,1\rangle}.
\end{eqnarray}
where $P_{\left|\psi\right\rangle}$ is the population in state $\left|\psi\right\rangle$ and $\Gamma_{\left|\psi\right\rangle}$ is the decay rate of $\left|\psi\right\rangle$ into the target state $\left|\xi,0\right\rangle$. Following the same procedure as for the calculation of steady state error, for a fixed ratio of coupling parameters, the expressions for $\Gamma$ can be expanded in a relevant small parameter to obtain the respective rates in a given regime. Here we report the value of $\Gamma_{C,0}$, calculated for $g_{r} = (3/4)g$,
\begin{eqnarray}
\Gamma_{C,0}  & = & \left\{
                            \begin{array}{cc}
                            \displaystyle 0.49 \left(\frac{g^{2}}{\kappa}\right);& \kappa \gg g\\
                            \\
                            \displaystyle\frac{\kappa}{2};& \kappa \ll g.
                            \end{array}
                        \right.
\label{Eq:AnalyticalTau}  
\end{eqnarray}
The above expressions lead to a net preparation rate, as
\begin{eqnarray}
    \widetilde{\tau}^{-1} & = & \Gamma_{\rm eff}= (\Gamma_{C,0}^{-1}|_{\kappa\ll g}+\Gamma_{C,0}^{-1}|_{\kappa\gg g})^{-1}\nonumber\\
    & \approx & \frac{\kappa g^{2}}{2 (g^{2} + \kappa^{2})}.
    \label{Eq:AnalyticalTaueff}  
\end{eqnarray}
Analytical estimate of preparation rate in Eq. (\ref{Eq:AnalyticalTaueff}) serves as an upper bound for the exact result obtained from the Liouvillian gap, i.e. $\Gamma_{\rm eff} \geq {\rm Re}[\Delta_{\mathcal{L}}]$. This is because the total convergence rate is determined by a series of processes that shuffle excitations between multiple decoupled subspaces. It is also worth noting that, to leading order, qubit decays do not affect the convergence time of the scheme as also confirmed by numerical optimization.
%
%
\section{Numerical Optimization}
%
At any instant in time during the operation of the stabilization protocol, the error from the target state is 
\begin{equation}
\epsilon(t) = 1 - {\rm Tr} \left[\rho(t) \mathbb{I}_{\rm res}\otimes|\xi\rangle\langle \xi|\right].
\end{equation}
This error can be decomposed into a static and dynamical error as 
\begin{equation}
\varepsilon(t) = \varepsilon_\infty + \tilde{\varepsilon}\exp\left(-t/\tau\right),
\end{equation}
where the dynamical error decays exponentially to zero as $t\to\infty$ with some characteristic lifetime $\tau$ leaving only the steady-state error $\varepsilon_\infty$. Both $\varepsilon_\infty$ and $\tau$ may be accurately calculated by time-domain simulations of the master equation. Alternately, both $\varepsilon_\infty$ and $\tau$ may be determined directly from the Liouvillian as $\epsilon_{\infty} = 1 - {\rm Tr} (\rho_{ss} \mathbb{I}_{\rm res}\otimes|\xi\rangle\langle \xi|)$ and $\tau = \Delta_{\mathcal{L}}^{-1} = -\textrm{Re}[\lambda_1]^{-1}$ respectively, where $\mathcal{L}\rho_{ss} =0$ and $\lambda_1$ is the lowest lying non-zero eigenvalue of the Liouvillian. We have confirmed that these two methods of calculating the performance metrics of our stabilization protocol are in excellent agreement, with relative errors on the order of $10^{-6}$ for both $\varepsilon_\infty$ and $\tau$.
%
%
\subsection{Robustness to imperfections and asymmetries}
%
\begin{figure}[t!]
\centering
\includegraphics[width=0.5\textwidth]{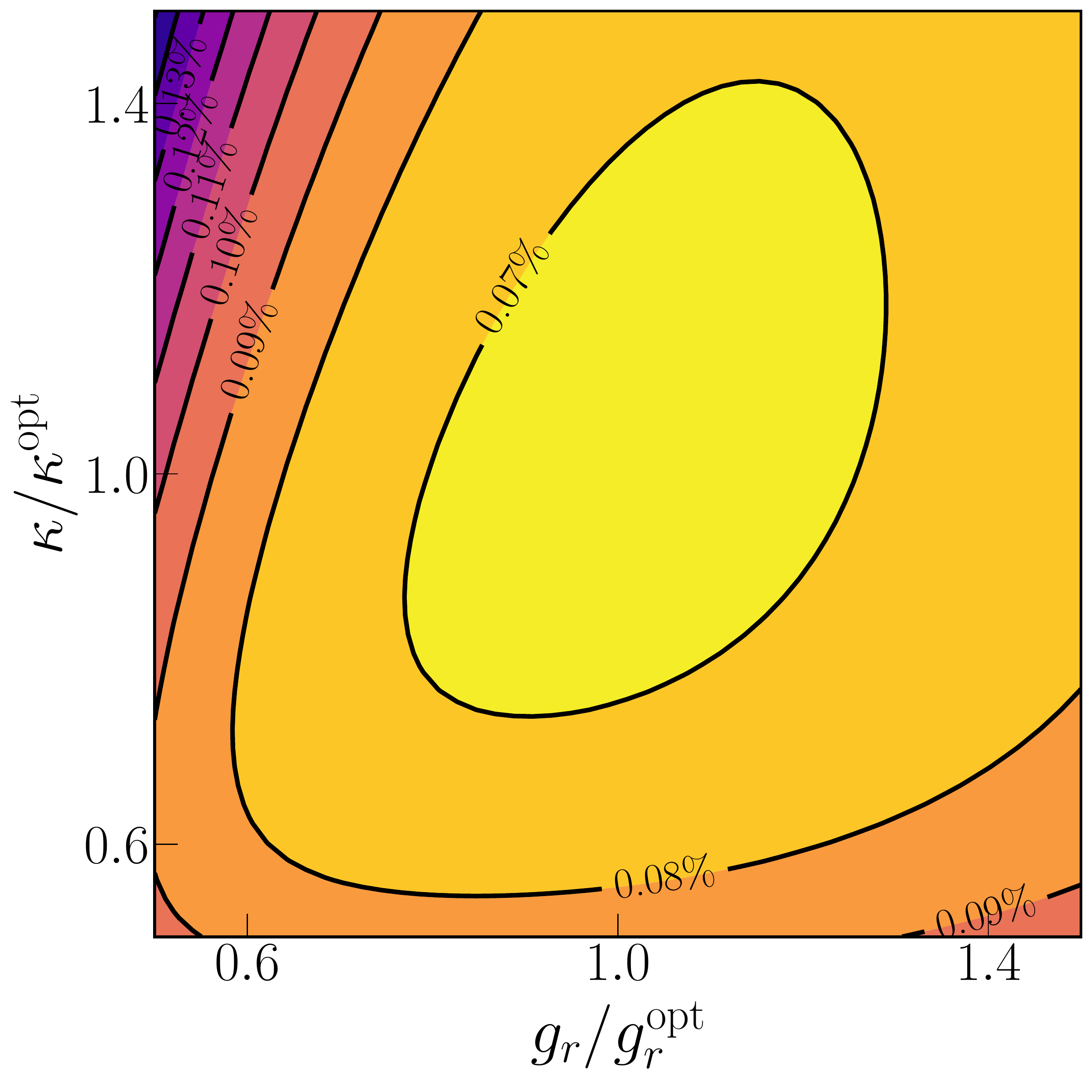}
\caption{Steady-state error as a function of the deviation of the resonance linewidth $\kappa$ and qubit-resonator coupling strength $g_r$ from their respective optimal values, $\kappa^{\rm opt}$ and $g_r^{\rm opt}$. This simulation is performed with $g = (2/3)\kappa^{\rm opt} = (4/3)g_r^{\rm opt} = 2\pi \times 50\;\textrm{MHz}$, and assuming $T_{1}= 100\; \mu {\rm s}, T_{2} = 200\; \mu {\rm s}$.} 
\label{Fig:Robust}
\end{figure}
\begin{figure}[t!]
\centering
\includegraphics[width=\textwidth]{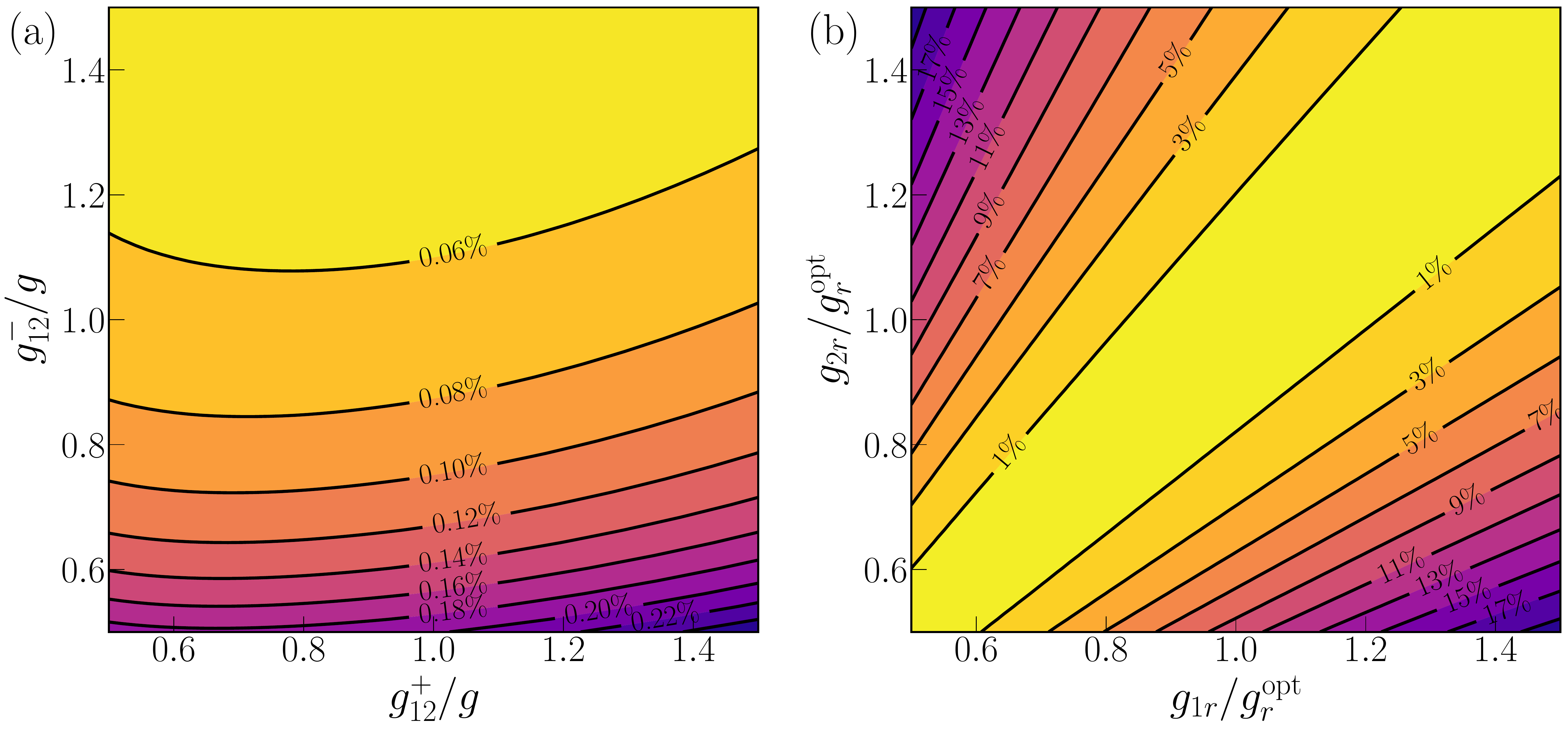}
\caption{(a) Steady-state error for asymmetric qubit-qubit couplings, calculated with $g = (2/3)\kappa^{\rm opt} =
    (4/3)g_{r}^{\rm opt} = 2\pi\times50\;\textrm{MHz}$. Even large asymmetries have a limited effect on the achievable fidelities, and can in fact
    provide slightly better performance when $g_{12}^{-} > g_{12}^{+}$. (b) Steady-state error for asymmetric
    qubit-resonator couplings, for the same parameter values. In contrast to the qubit-qubit couplings, asymmetry in
    the qubit-resonator couplings has a detrimental effect on the achievable fidelity. This can be remedied by leveraging the tunable strength of the parametric couplings.}
\label{Fig:Asymmetry}
\end{figure}
\begin{figure}[t!]
\centering
\includegraphics[width=\textwidth]{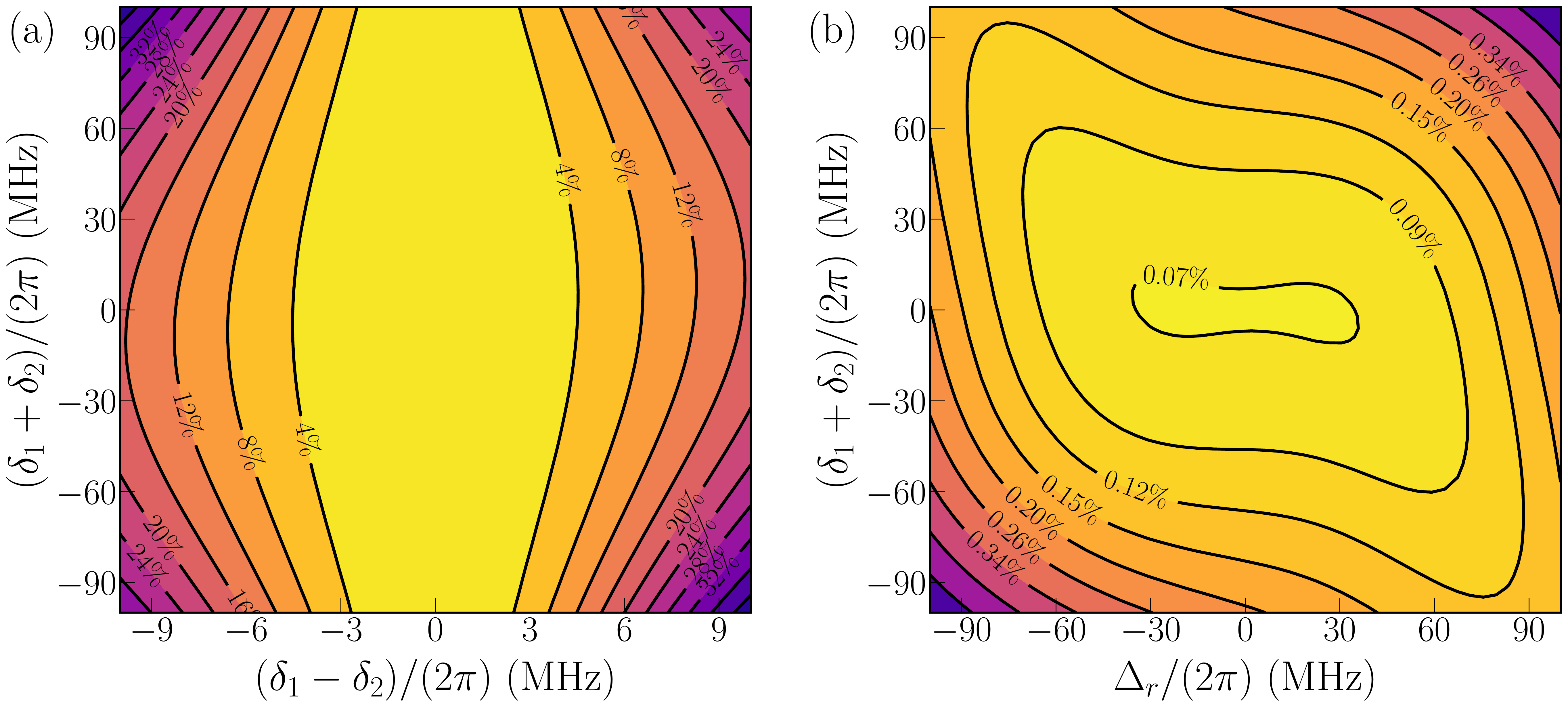}
\caption{(a) Steady-state error when preparing an odd-parity state in the presence of qubit detunings, calculated with $g = (2/3)\kappa^{\rm opt} = (4/3)g_r^{\rm opt} = 2\pi \times 50\;\textrm{MHz}$, and assuming $T_{1}= 100\; \mu {\rm s}, T_{2} = 200\; \mu {\rm s}$. When $\delta_1 - \delta_2 \neq 0$, there is spurious coupling between $\ket{S}$ and $\ket{T}$ which limits the achievable fidelity. The scheme is mostly insensitive to $\delta_1 + \delta_2 \neq 0$ when stabilizing an odd-parity state. When stabilizing an even-parity state, the sum of $\delta_1$ and $\delta_2$ leads to spurious coupling rather than the difference. (b) Steady-state error for nonzero qubit and cavity detunings when preparing an odd-parity state, for the same parameters and with $\delta_1 - \delta_2 = 0$, demonstrating the slight performance increase available by detuning the resonator.}
\label{Fig:Detuning}
\end{figure}
The scaling of performance metrics of our stabilization protocol as a function of $g$ is demonstrated in Figure 2 in the main text. Specifically, the figure shows the effect of varying $g$ while all other parameters are held constant. Increasing $g$ improves the fidelity and the stabilization rate, both of which eventually saturate until the qubit-resonator coupling $g_{r}$ and resonator linewidth $\kappa$ are also increased. For a specified $g$, both $g_r$ and $\kappa$ have optimal values determined as $\kappa^{\rm opt} = 2g_r^{\rm opt} = (3/2)g$. Figure \ref{Fig:Robust} shows the robustness of the stabilization protocol to deviations of $\kappa$ and $g_r$ from their respective optimal values. Large variations in either or both parameters have a limited effect on the achievable fidelities of the stabilization protocol. 
\par
Figure \ref{Fig:Asymmetry} shows the effect of asymmetries between either the two qubit-qubit couplings $g_{12}^{+}$ and $g_{12}^{-}$ or between the two qubit-resonator couplings $g_{1r}$ and $g_{2r}$. Of the two possible asymmetries, the case where $g_{1r} \neq g_{2r}$ has a much larger negative impact on the achievable fidelities. This is expected as these couplings are used exclusively to engineer the collapse operator acting on the two-qubit subspace. It is worth noting that since parametric coupling strengths can be tuned with pump amplitudes, in practice the asymmetry between such couplings can be mitigated.
\par
The effect of detunings on the achievable fidelities is shown in Figure \ref{Fig:Detuning}. When the qubit detunings $\delta_1$ and $\delta_2$ are nonzero, there may be spurious coupling introduced between the target state $\ket{\xi}$ and $\ket{\bar{\xi}}$. This coupling is introduced if either $\delta_1 + \delta_2 \neq 0$ or $\delta_1 - \delta_2 \neq 0$, when stabilizing an even- or odd-parity state respectively. It is therefore natural to parameterize the qubit detunings in terms of the sum and difference $\delta_1 \pm \delta_2$. The protocol is sensitive to the parameter which leads to the spurious coupling, and mostly insensitive to the conjugate parameter. In practice this is not a serious limitation, given the precise frequency control possible with microwave electronics.
\par
Figure \ref{Fig:Detuning} also shows the effect of detuning of the resonator. The scheme is weakly sensitive to this detuning, though it demonstrates a slight improvement of steady-state fidelity with some amount of detuning $0 < |\Delta_r| < g_r$. By detuning the resonator, the Rabi contrast between the states $\ket{C,0}$ and $\ket{\bar{\xi},1}$ is reduced which ultimately leads to a slightly increased stabilization rate and thus steady-state fidelity.
%
\subsection{Effect of counter-rotating terms}
%
\begin{figure}[t!]
\centering
\includegraphics[width=0.6\textwidth]{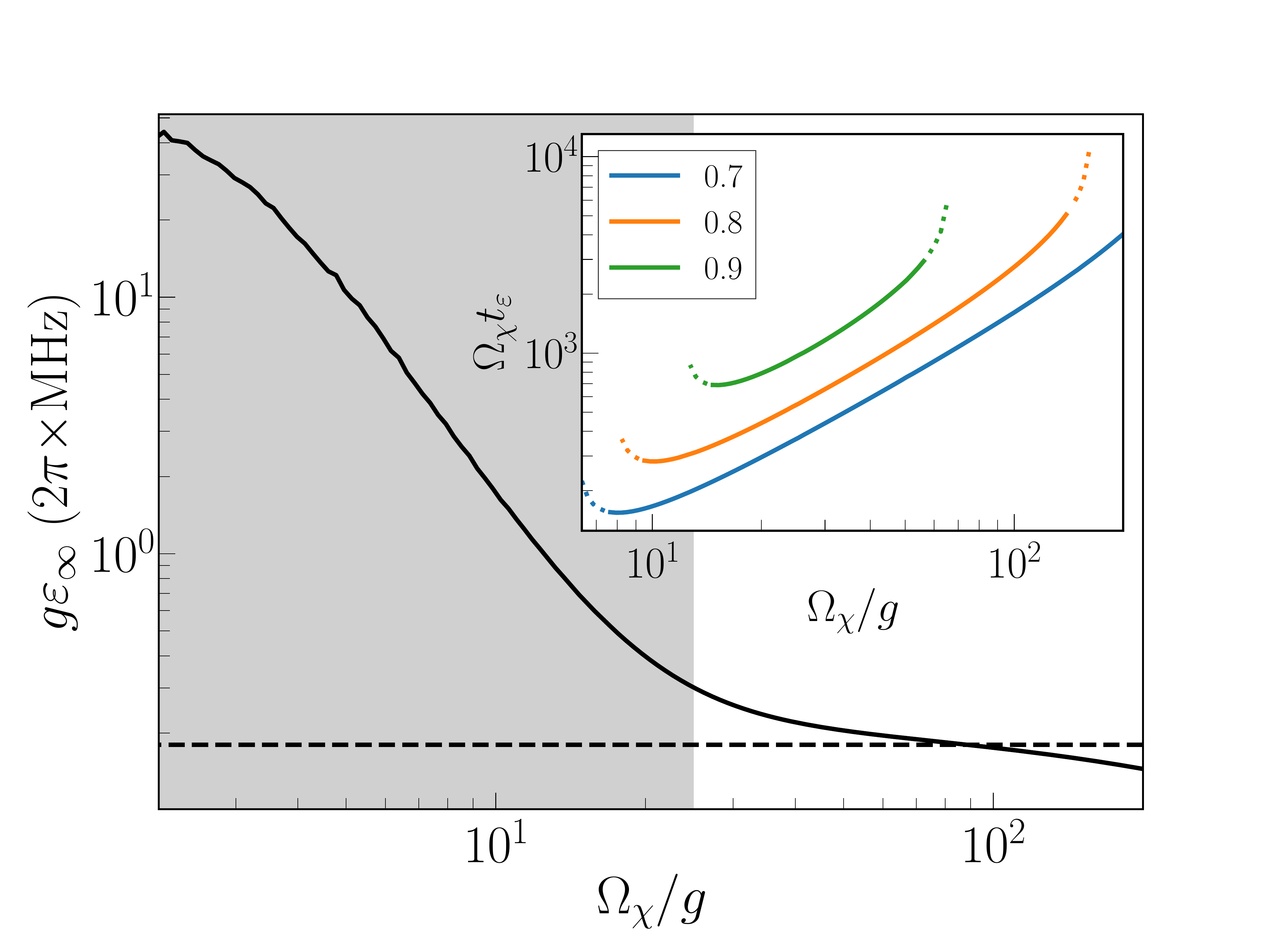}
\caption{Scaling of steady-state error for even-parity Bell state preparation, calculated via master equation 
simulations including the counter-rotating terms, with $\chi_{1,2} = 2\pi\times10\;\textrm{MHz}, T_{1}= 100\; \mu {\rm s}, T_{2} = 200\; \mu {\rm s}$. The gray region corresponds to the parameter regime where counter-rotating terms induce a 
fidelity vs. speed tradeoff. The downturn of the curve in the region of large $\Omega_\chi/g$ is due to $g$ 
becoming comparable to $\gamma_1$. The dashed line is an approximation of the steady-state error derived from the
Liouvillian [Eq.~(\ref{Eq:AnalyticalErrorOpt})], in the limit $\gamma_1/g\ll1$. (Inset) Time $t_{\varepsilon}$ taken to reach a given fidelity threshold, for the same $\Omega_\chi = 2\pi\times20\;{\rm MHz}$. When the threshold is near the maximum achievable 
fidelity for a given set of parameters, the nature of the scaling changes. When the threshold is sufficiently less than the achievable fidelity, then the threshold crossing will occur in the region where the dynamical error dominates and the error is decreasing exponentially. If however the threshold is close to the maximum fidelity, then the crossing will occur when the static error is comparable to or larger than the dynamic error and thus the absolute error is falling at a decreasing rate. The regions in which this behaviour is observed are the dotted portions of the curves, caused either by $g$ being too large and thus the leakage dominating, or by $g$ being comparable to $\gamma_1$.}
\label{Fig:Scaling}%
\end{figure}
The Hamiltonian considered in the main text ignores all non-resonant terms. Fig.~\ref{Fig:Scaling} shows the performance metrics for even-parity Bell state stabilization, including all the off-resonant terms of the form reported in Eq.~(\ref{eqn:HNoRwa}). We have found that the qualitative effect of these terms can be captured by introducing a single term describing $\chi$-dependent leakage out of the target state,
\begin{eqnarray}
	\mathcal{H}_{\rm CR} \approx g e^{i\phi_{12}^{\pm}}\left(e^{i 2\Omega_\chi t}| \bar{\xi},0 \rangle\langle \xi, 0| \right) + h.c.
\end{eqnarray}
where $\Omega_\chi = \chi_1 \pm \chi_2$. The phase on the prefactor and the sign in the definition of $\Omega_\chi$ is determined by the pump frequencies; when preparing an even parity Bell state $+$ is selected, and $-$ for odd. 
\par
Recall that when considering the resonant Hamiltonian alone, increasing $g$ does not reduce speed or fidelity as shown in Figure~2 of the main text. Instead, both the speed and achievable steady-state fidelity continue to improve, until they saturate due to processes mediated by $g_{r}$ or $\kappa$ being the bottleneck. As is clear from  Fig.~\ref{Fig:Scaling}, the scaling in the presence of $\chi$-dependent leakage is different, since the effect of the leakage term does not saturate with increasing $g$. Therefore, in contrast to the resonant case where it is only necessary to optimize $g_r$ and $\kappa$, introduction of the leakage necessitates setting $g$ to some optimal value $g^{\rm opt}$ found numerically for a given $\Omega_\chi$. Thus the parameter space can now be split into two qualitatively different regimes of operation depending on the ratio $g^{\rm opt}/g$. When $g^{\rm opt}/g \gg 1$ the leakage term has little effect and the scaling behavior of the resonant Hamiltonian is recovered, where $\varepsilon_\infty$ and $\tau$ both scale as $1/g$ assuming $g_r$ and $\kappa$ scale optimally with $g$. When $g^{\rm opt}/g \ll 1$, the achievable steady state fidelity from the resonant Hamiltonian saturates and the scaling of $\varepsilon_\infty$ is dominated by the increasingly strong leakage term. In this regime, the ratio $\varepsilon/\tau$ is no longer constant. These two regimes are demarcated in Figure \ref{Fig:Scaling}.
%
\section{Coupler-induced qubit decoherence}
%
For the theoretical results presented in the manuscript, we considered an ideal parametric coupler which induces no additional decay channel on the qubit subsystem. In this section we present some empirical estimates of qubit decoherence induced due to coupler architecture presented in Fig. 4 of the main manuscript. We first estimate qubit relaxation due to modification of the effective input impedance, seen by the qubit circuit, by the input flux line. The resultant lifetime $T_{1j}$ of qubit $j$ limited by coupler impedance can be expressed as
\begin{equation}
T_{1j} = \frac{L_j}{\Re{Z_{\rm in}[\omega_j]}},
\end{equation}
where $\Re{Z_{\rm in}[\omega_j]}$ is the environment impedance seen by the qubit $j$ at its resonant frequency. As a first approximation, we can assume that the flux line is a transmission line with characteristic impedance  $Z_0=50\Omega$, that terminates into an inductance $L_0$ sharing a mutual inductance $M$ with the SQUID loop. We can then compute $Z_{in}$ as
\begin{equation}
Z_{in}[\omega]=i\omega(L_{sq}-M)+\frac{i\omega M(i\omega(L_0-M)+Z_0)}{i\omega L_0 + Z_0},
\end{equation}
where $L_{sq}$ denotes the inductance of the SQUID loop. Therefore, the qubit lifetime in the limit $L_j \gg L_0,L_{sq}$ is
\begin{equation}
T_{1j} = \frac{L_j(\omega_j^2L_0^2+Z_0^2)}{\omega_0^2M^2Z_0}.
\end{equation}
For typical values of the parameters in circuit-QED parameters, $L_j=20\;{\rm nH}$, $\omega_j=2\pi\times 5\;{\rm GHz}$, $L_0=0.1\;{\rm nH}$, $M=2\;{\rm pH}$, we obtain the estimated $T_{1j}\approx 240\;\mu s$. In general we expect that the environment impedance might be complex and modified by resonances in the control line, which could further limit qubit lifetime via Purcell decay. This can be mitigated  by placing a notch filter centered at the qubit frequencies into the coupler flux line \cite{Lu2017}. Designing such a filter should be straightforward as long as the frequencies of the parametric pumps are well separated from the qubit and resonator frequencies.
\begin{figure}[t!]
\centering
\includegraphics[width=0.8\textwidth]{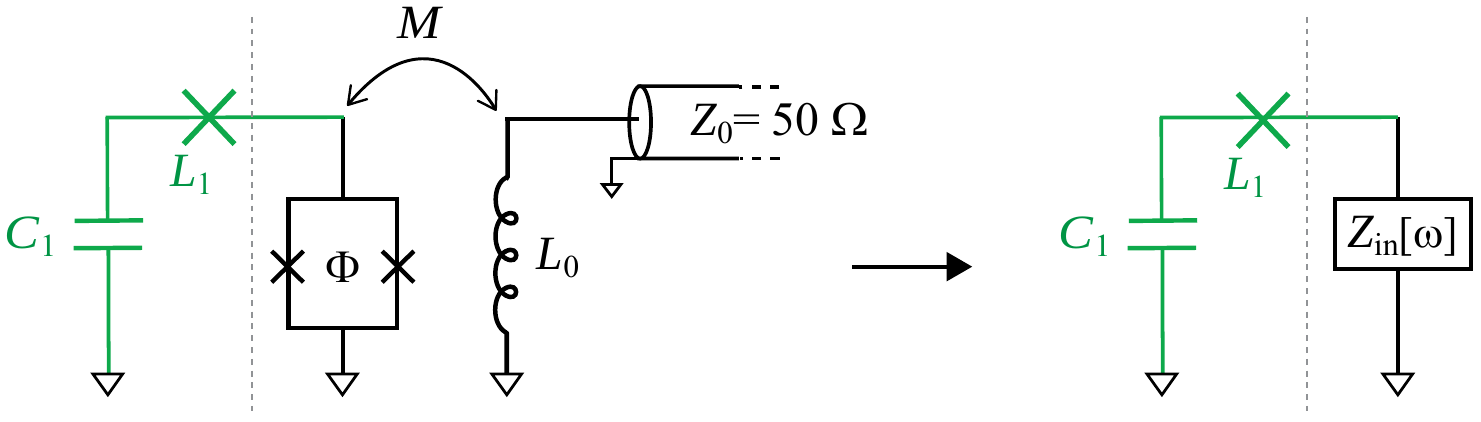}
\caption{(Left panel) Circuit schematic showing the inductive coupling of the qubit to the flux line, mediated through the coupler. (Right panel) Series LCR circuit obtained using the effective input impedance seen by the qubit due to the flux coupling line.}
\label{Fig:CouplerDecoherence}%
\end{figure}
\par
The flux noise seen by the coupler can also lead to frequency jitter in the qubit, leading to additional dephasing. To estimate this we consider $1/f$-type flux noise spectrum
\begin{eqnarray}
S_{\Phi\Phi} [\omega_{j}] = \frac{A^{2}}{\omega_{j}} (2 \pi \times 1 \,{\rm Hz}),
\end{eqnarray}
for which the dephasing time can be estimated as
\begin{eqnarray}
    T_{2j}^{-1} \approx \Delta \omega_{j} & = & \left|\frac{\partial \omega_{j}}{\partial \Phi}\right|\Delta \Phi= \frac{\omega_{j}}{2(L_{j} + L_{sq}(\Phi))} \left(\frac{\partial L_{sq} (\Phi)}{\partial \Phi}\right) \Delta \Phi \nonumber\\
    & \approx & \left(\frac{\pi\omega_{j}}{2}\right)\frac{L_{sq} (\Phi)}{L_{j} + L_{sq} (\Phi)} \tan\left(\frac{\pi \Phi}{\Phi_{0}}\right) \sec\left(\frac{\pi \Phi}{\Phi_{0}}\right) \Delta \Phi
\end{eqnarray}
with average flux fluctuation 
\begin{eqnarray}
    \Delta \Phi = \left(\int_{\omega_{\rm min}}^{\omega_{\rm max}} d\omega \; S_{\Phi\Phi} [\omega]\right)^{1/2}.
\end{eqnarray}
The above expression is logarithmic in the upper and lower cut-off frequencies, with the latter typically set by the measurement bandwidth and ensuring the convergence of the integral \cite{Ithier2005}. For typical flux noise amplitudes, $A = 1-2\; {\mu \Phi_{0}}$, and circuit parameters (see Fig. 4 of the main manuscript), this gives an estimated  $T_{2j}^{*} \geqslant 50 \; {\mu}s$. A gradiometric design for the coupler can be used to further insulate the junction from the flux noise due to coupling inductance.
%
%

%